\documentclass[journal,twocolumn]{IEEEtran}
\usepackage{epsfig,makeidx,color}
\usepackage{amsmath,amssymb,bbm}
\usepackage{cite,graphicx,lipsum}
\usepackage{enumerate}
\usepackage[switch,pagewise]{lineno}
\usepackage{hyperref}
\hypersetup{
        colorlinks = true,
        citecolor=blue,
}
\def\cQ{{\cal Q}}

\def\rH{{\rm H}}
\def\rT{{\rm T}}

\def\uP{{\mathbb P}}

\def\uE{{\mathbb E}}

\newtheorem{mylemma}{\bf Lemma} 

\def\deft{ \buildrel \triangle \over = }

\def\be{ \begin{equation} }
\def\ee{ \end{equation} }
\def\bea{ \begin{eqnarray} }
\def\eea{ \end{eqnarray} }

\def\by{{\bf y}}

\def\bs{{\bf s}}
\def\ba{{\bf a}}
\def\br{{\bf r}}

\def\bn{{\bf n}}
\def\bh{{\bf h}}

\def\bE{{\bf E}}

\def\bI{{\bf I}}

\def\bR{{\bf R}}

\def\bX{{\bf X}}

\def\b0{{\bf 0}}

\def\cC{{\cal C}}

\def\cQ{{\cal Q}}

\def\cN{{\cal N}}

\ifCLASSOPTIONonecolumn
  \newcommand{\figwidth}{0.40\columnwidth}
\else
  \newcommand{\figwidth}{0.95\columnwidth}
\fi

\begin{document}

\title{On Error Rate Analysis for URLLC 
over Multiple Fading Channels}

\author{Jinho Choi\\
\thanks{The author is with
the School of Information Technology,
Deakin University, Geelong, VIC 3220, Australia
(e-mail: jinho.choi@deakin.edu.au).}}


\maketitle
\begin{abstract}
In this paper, we study
ultra-reliable and low-latency communication (URLLC)
under fading using multiple frequency or time bins.
We investigate an approach to find an upper-bound on
the packet error rate when a finite-length code is used.
From simulation results, we find that the bound is reasonably tight 
for wide ranges of signal-to-noise ratio (SNR) and 
the number of multiple bins. Thus, the derived
bound can be used to 
determine key parameters to guarantee the performance
of URLLC in terms of the packet error rate.
\end{abstract}


\ifCLASSOPTIONonecolumn
\baselineskip 24pt
\fi

\section{Introduction}

Ultra-reliable and low-latency communication (URLLC)
has been considered
for a number of real-time applications
such as factory automation, autonomous driving, and remote surgery.
In 5th generation (5G) cellular systems,
URLLC is to be supported \cite{Soldani18} \cite{Bennis18}.
For URLLC,
in \cite{Anand18}, resource allocation
and hybrid automatic repeat request (HARQ) schemes are 
investigated.
In \cite{Choi19}, the notion of effective 
bandwidth \cite{Chang95} is studied in order to
guarantee a certain latency with 
quality of service (QoS) exponent.
URLLC in machine-type communication (MTC) \cite{Bockelmann16}
is studied with random access in \cite{Malak19}.

In order to perform reliable transmissions, 
channel coding can be employed.
Since the length of packets can be short
it is necessary to consider finite-length codes
and understand their impact on the performance
\cite{Polyanskiy10IT}.
In \cite{Durisi16}, URLLC is considered with finite-length codes.
In \cite{Shirvan},
various existing channel codes are studied and compared with
the theoretical performance obtained in \cite{Polyanskiy10IT}.

As in \cite{Wolf17} \cite{Rao18},
multi-connectivity can be used to provide a diversity gain,
which improves the reliability in transmissions over fading.
In this paper, we also consider multichannel transmissions
to exploit a high diversity gain 
so that one transmission (through multiple channels) is sufficient
as its packet error rate can be low
without re-transmission in URLLC.
In order to guarantee a sufficient low packet error
rate, it is necessary to decide key parameters in advance,
which requires a good prediction of the 
packet error rate.
To this end, in this paper, we focus on the derivation
of an upper-bound on the packet error probability
when finite-length codes are used in URLLC.

\subsubsection*{Notation}
Matrices and vectors are denoted by upper- and lower-case
boldface letters, respectively.
The superscript $\rT$ denotes the transpose.
The 2-norm is denoted by $||\cdot||$.
For a matrix $\bX$, $[\bX]_{m,n}$ represents the $(m,n)$th element.
$\uE[\cdot]$ and ${\rm Var}(\cdot)$
denote the statistical expectation and variance, respectively.
$\cC \cN(\ba, \bR)$
represents the distribution of
a circularly symmetric complex Gaussian
(CSCG) random vector with mean vector $\ba$ and
covariance matrix $\bR$.

\section{System Model}

Suppose that we have a set of $L$ frequency bins or blocks
for multi-connectivity in URLLC \cite{Wolf17}.
A transmitter is to transmit the same coded packet
through $L$ blocks.
Then, the received signal at a receiver through 
the $l$th block is given by
\be
\br_l = h_{l} \bs + 
\bn_l, \ l \in \{1,\ldots, L\},
\ee
where $h_{l}$ represents
the channel coefficient from the transmitter to the receiver
through the $l$th block,
$\bs$ is a coded packet,
and $\bn_l \sim \cC \cN(\b0, N_0 \bI)$ is the background noise vector.

To decode the signal, we can consider the maximal
ratio combining (MRC) \cite{BiglieriBook} \cite{ChoiJBook2}
as follows:
\begin{align}
\by & = \sum_{l} h_{l}^* \br_l \cr
& = \sum_{l} |h_{l}|^2 \bs + h_l^* \bn_l.
	\label{EQ:byk}
\end{align}
Then, the instantaneous 
signal-to-noise ratio (SNR) after MRC, which is
referred to as MRC-SNR for convenience, is given by
\be
\rho = \frac{ ||\bh||^2 P}{N_0},
\ee
where $\bh = [h_1 \ \ldots  \ h_L]^\rT$ and
$\bE[\bs \bs^\rH] = P \bI$. Here, $P$ represents
the signal transmit power and $\frac{P}{N_0}$ is referred to as the SNR.
For a reliable communication, a high diversity gain
with a sufficient SNR is required.
In addition, if the packet error rate
is sufficiently low with a large $L$, no re-transmission
might be required, which can result in low-latency communication.
Thus, it is expected to predict  
the packet error rate in terms of $L$ and SNR 
in URLLC.

Note that the outage probability of MRC is well-known
\cite{ProakisBook}.
However, when finite-length codes are used \cite{Polyanskiy10IT},
the packet error rate cannot be directly expressed 
by the outage probability.

\section{Error Probability Analysis}

In URLLC, it is necessary to decide key parameters (e.g.,
the signal transmit power, $P$, and the number of blocks, $L$)
to provide a certain guaranteed performance. For example,
we can consider the packet error rate.
In this section, we find a closed-form expression
for the packet error rate that helps decide the values of key
parameters
when finite-length codes are used.

\subsection{Error Probability of Finite Length Codes}

For a given $\rho$, according to \cite{Polyanskiy10IT} \cite{Tan15},
the achievable rate
(for complex Gaussian channel \cite{Durisi16})
is given by
\be
R^* (n, \epsilon) \approx \log_2 (1 + \rho)
- \sqrt\frac{ V (\rho)}{n} \cQ^{-1} (\epsilon) + \frac{\log_2 n}{2 n},
        \label{EQ:R_PPV}
\ee
where $V(\rho)$ is the channel dispersion that is given by
\begin{align}
V(\rho)
= \frac{\rho(2 + \rho)}{(1+ \rho)^2} (\log_2 e)^2 ,
\end{align}
$n$ is the length of codeword when a codeword
is transmitted within a block,
and $\epsilon$ is the nominal\footnote{This becomes the error probability
when $\rho$ is fixed. However, if $\rho$ is
a random variable (due to fading), we also need to take
into account the outage probability to find
the error probability.
From this reason, it is referred to
as the \emph{nominal} error probability.}
error probability.
It can be shown that $\bar V > V(\rho)$, where
$\bar V = \frac{1}{(\ln 2)^2} \approx 2.0814$ \cite{Choi19}.
Thus, ignoring the term of
$O\left(\frac{\log_2 n}{n}\right)$,
a lower-bound on the achievable rate can be obtained as follows:
\be
{\underline R} (\rho, n, \epsilon)
=
\log_2 (1 + \rho)
- \sqrt\frac{\bar V}{n} \cQ^{-1} (\epsilon),
        \label{EQ:LB}
\ee
which might be tight for a sufficiently high 
MRC-SNR, $\rho$,
because $V(\rho) \to \bar V$ as $\rho \to \infty$.
The lower-bound in \eqref{EQ:LB} allows tractable analysis,
since the terms of $\epsilon$ and $\rho$ are decoupled.

\begin{mylemma}
With a code rate $R$, the probability of unsuccessful decoding
is upper-bounded as
\begin{align}
\uP_{\rm err} 
\le \epsilon + (1- \epsilon)\Pr(\rho <  \tau (\epsilon)),
	\label{EQ:UP_l}
\end{align}
where 
\be
\tau (\epsilon)
= 2^{R + \sqrt{\frac{\bar V}{n}} \cQ^{-1} (\epsilon)} - 1.
\ee
\end{mylemma}
\begin{IEEEproof}
For convenience, let $\tau = \tau (\epsilon)$.
From \eqref{EQ:LB}, the probability that 
the achievable rate is lower than $R$ is upper-bounded by
$\Pr(\rho <  \tau)$.
To find an upper-bound on 
the probability of unsuccessful decoding
we can assume that the decoding fails
if $\rho < \tau$ with probability 1.
Then, we have
\begin{align}
\uP_{\rm err} \le \Pr({\rm err} \,|\, \rho \ge \tau)
\Pr(\rho \ge \tau) + \Pr(\rho <  \tau),
\end{align}
where $\Pr({\rm err} \,|\, \rho \ge \tau)$
is the conditional probability of decoding error 
for given $\rho \ge \tau$.
Thus, $\Pr({\rm err} \,|\, \rho \ge \tau)$ is upper-bounded by
$\epsilon$, and we have
$$
\uP_{\rm err} \le \epsilon \Pr(\rho \ge \tau) + \Pr(\rho <  \tau),
$$
which becomes \eqref{EQ:UP_l}.
\end{IEEEproof}

By taking $\epsilon$ as a parameter to minimize the upper-bound
in \eqref{EQ:UP_l},
we can have the following tight upper-bound:
\begin{align}
\uP_{\rm err} \le \bar \uP_{\rm err} 
\deft \min_{0 \le \epsilon \le 1} \epsilon + (1- \epsilon)
\Pr(\rho <  \tau (\epsilon)).
	\label{EQ:bP}
\end{align}
In \eqref{EQ:bP},
$\Pr(\rho <  \tau (\epsilon))$ is referred to
as the outage probability.

Note that if $n \to \infty$
and a capacity achieving code is used,
the packet error rate can be simply expressed by
the outage probability, i.e.,
\begin{align*}
\uP_{\rm err} 
= \Pr (\log_2 (1+ \rho) < R) 
= \Pr \left(\rho < \tau\right),
\end{align*}
where $\tau = 2^{R} - 1$.

\subsection{Outage Probability}

In this subsection, 
we consider a closed-form expression for the outage probability.

We assume independent Rayleigh fading channels and
\be
\uE[h_{l} h_{l^\prime}^*] = \sigma_h^2 \delta_{l,l^\prime}.
\ee
Thus, $|h_{l}|^2$
has the following exponential distribution:
\be
|h_{l}|^2 \sim {\rm Exp}(\sigma_h^2) = \frac{1}{\sigma_h^2} \exp
\left(
- \frac{|h_{l}|^2}{\sigma_h^2}
\right).
	\label{EQ:hpdf}
\ee
From \eqref{EQ:hpdf}, we can show that
$$
\frac{||\bh||^2 }{\sigma_h^2}
= \frac{\chi_{2 L}^2}{2},
$$
where $\chi_n^2$ represents a chi-squared random
variable with $n$ degrees of freedom.
For convenience, let
\be
Z_L = \frac{\chi_{2L}^2}{2L}.
\ee
The cdf of $\rho$ is given by
\begin{align}
\Pr(\rho < \tau)
= \Pr \left( Z_{L} < \frac{\tau}{\beta} \right) 
= \frac{\gamma \left(L, \frac{\tau L}{\beta} \right)}{(L-1)!},
	\label{EQ:Prho}
\end{align}
where $\gamma (s,x)
= \int_0^x t^{s-1} e^{-t} dt$ is the lower incomplete gamma function
and
\be
\beta  =  \frac{LP \sigma_h^2}{N_0}.
\ee

In order to have a tractable expression for
\eqref{EQ:Prho}, an upper-bound
on the tail probability of $Z_d$ can be considered.
Using the Chernoff bound \cite{Mitz05},
it can be shown that
\begin{align}
\Pr(Z_L < z)
& \le  \uE[ e^{-t (Z_L - z)}] \cr
& = e^{2 L tz} \left( \frac{1}{1+ 2 t} \right)^L \cr
& = \left(  \frac{e^{2t z}}{1+ 2 t} \right)^L, \ t \ge 0.
\end{align}
Letting
$z = \frac{1}{1 + 2t}$,
we have
\be
\Pr(Z_L < z) \le U_L (z) \deft (z e^{1-z} )^L, \ z \in [0, 1),
	\label{EQ:CB}
\ee
which is reasonably tight.
Note that $z e^{1-z}$ increases in $z \in [0, 1)$. Thus,
the upper-bound in  \eqref{EQ:CB} increases with $z$.

However, if a low outage probability is considered
with $z \to 0$ for reliable communication
in URLLC, the upper-bound may not be satisfactory.
For a tight bound when $z \to 0$, 
a term can be introduced.
In particular, we replace $z$ with $c_L z$ in \eqref{EQ:CB},
where $c_L$ is the correction term,
and let
\be
B_L (z) = (c_L z e^{1-z c_L})^L.
\ee
The correction term $c_L$ can be decided to 
satisfy
\be
\lim_{z \to 0} \frac{B_L (z)}{\Pr(Z_L \le z)} = 1
	\label{EQ:lim1}
\ee
so that $B_L(z)$ can approach the
outage probability, $\Pr(Z_L \le z)$, when $z \to 0$.
Since \cite{AbramowitzBook}
\be
\gamma(s,x) = x^s \sum_{k=0}^\infty \frac{(-x)^k}{k! (s+k)},
\ee
from \eqref{EQ:Prho},
we have
\be
\Pr(Z_L \le z) = \frac{1}{L!} ((Lz)^L + O(z^{L+1}) ).
	\label{EQ:A1}
\ee
In addition, 
\be
B_L (z) = (c_L  z e)^L + O(z^{L+1}).
	\label{EQ:A2}
\ee
Thus, to satisfy \eqref{EQ:lim1},
from \eqref{EQ:A1} and \eqref{EQ:A2}
we have
\be
c_L = L e^{-1} \left( L! \right)^{- \frac{1}{L}}.
\ee
Since $n! > e \left(\frac{n}{e}\right)^n$, 
it can be shown that
\be
c_L < e^{-\frac{1}{L}} \le 1.
\ee
From this and the fact that $U_L(z)$ increases with $z$, 
it follows that
$$
B_L (z) \le U_L(z), \ z \in [0, 1).
$$

\begin{figure}[thb]
\begin{center}
\includegraphics[width=\figwidth]{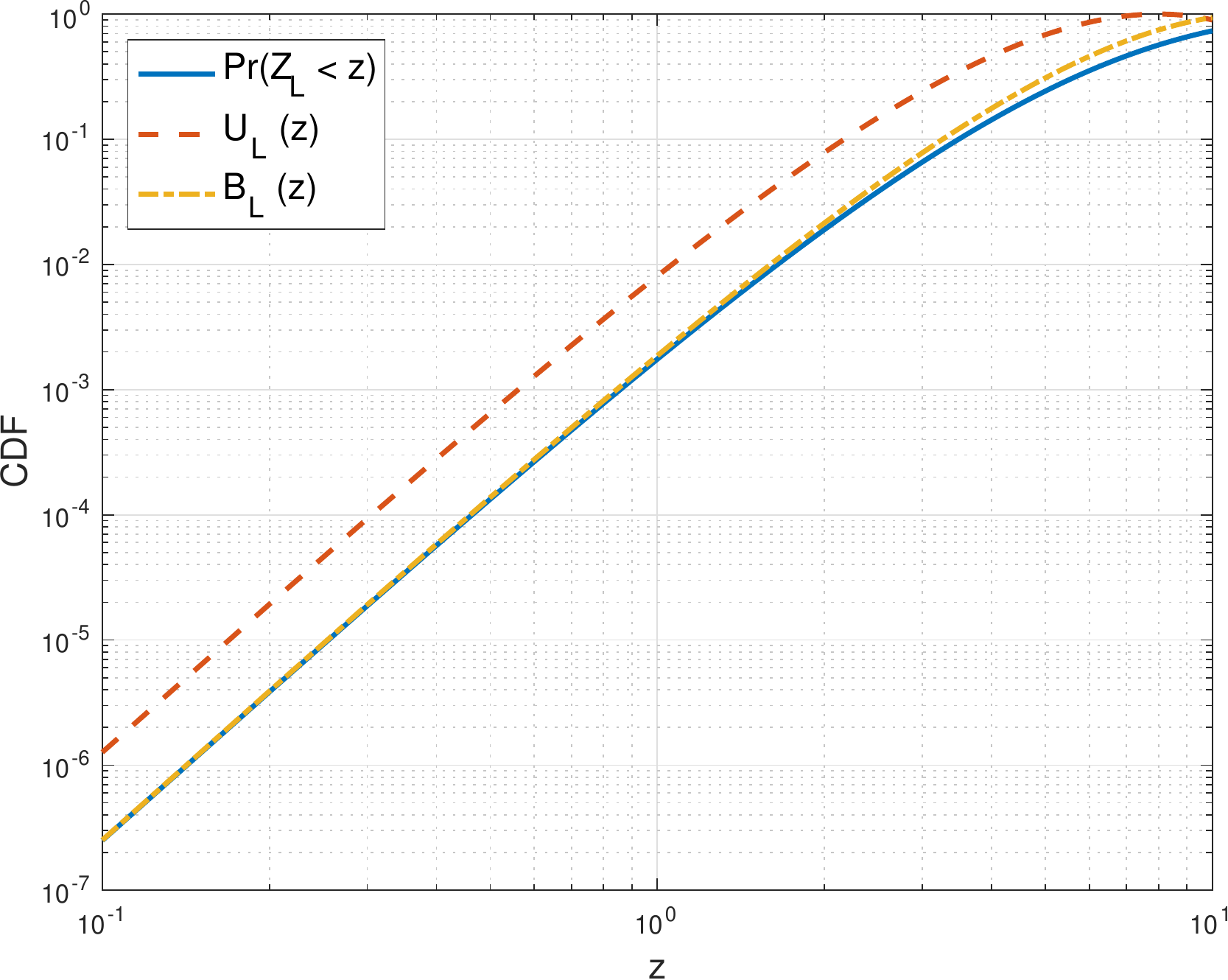}
\end{center}
\caption{Tail probability of $Z_L$ with $L = 4$: $\Pr(Z_L < z)$ with
$U_L(z)$ and $B_L (z)$.}
        \label{Fig:tails}
\end{figure}

Consequently, we can see that 
$B_L (z)$ can be tighter
than the upper-bound in \eqref{EQ:CB}
(i.e., $U_L(z)$)
and $B_L (z)$ approaches
$\Pr(Z_L < z)$ as $z \to 0$.
However, we are unable to prove that 
$B_L (z)$ is an upper-bound on 
$\Pr(Z_L < z)$, although 
it seems that
$B_L (z)$ is an upper-bound\footnote{In the rest of the paper, 
we conjecture that it is an upper-bound.} 
based on numerical results as shown in
Fig.~\ref{Fig:tails}.

With $B_L(z)$, \eqref{EQ:bP} is modified as follows:
\begin{align}
\bar \uP_{{\rm err}, B}
= \min_{0 \le \epsilon \le 1} \epsilon + (1- \epsilon)
B_L \left( \frac{\tau (\epsilon)}{\beta} \right).
	\label{EQ:bPB}
\end{align}
Since $B_L(z)$ is a closed-form expression,
a tight upper-bound can be found using a one-dimensional
numerical search algorithm.

\section{Simulation Results}

In this section, we present simulation results
under independent Rayleigh
fading channels with $\sigma_h^2 = 1$.
The bound in \eqref{EQ:bPB} is used to predict
the performance in terms of the packet error rate.
In URLLC, the packet error rate is to be $10^{-3} - 10^{-5}$
\cite{Simsek17} \cite{Bennis18}.

\begin{figure}[thb]
\begin{center}
\includegraphics[width=\figwidth]{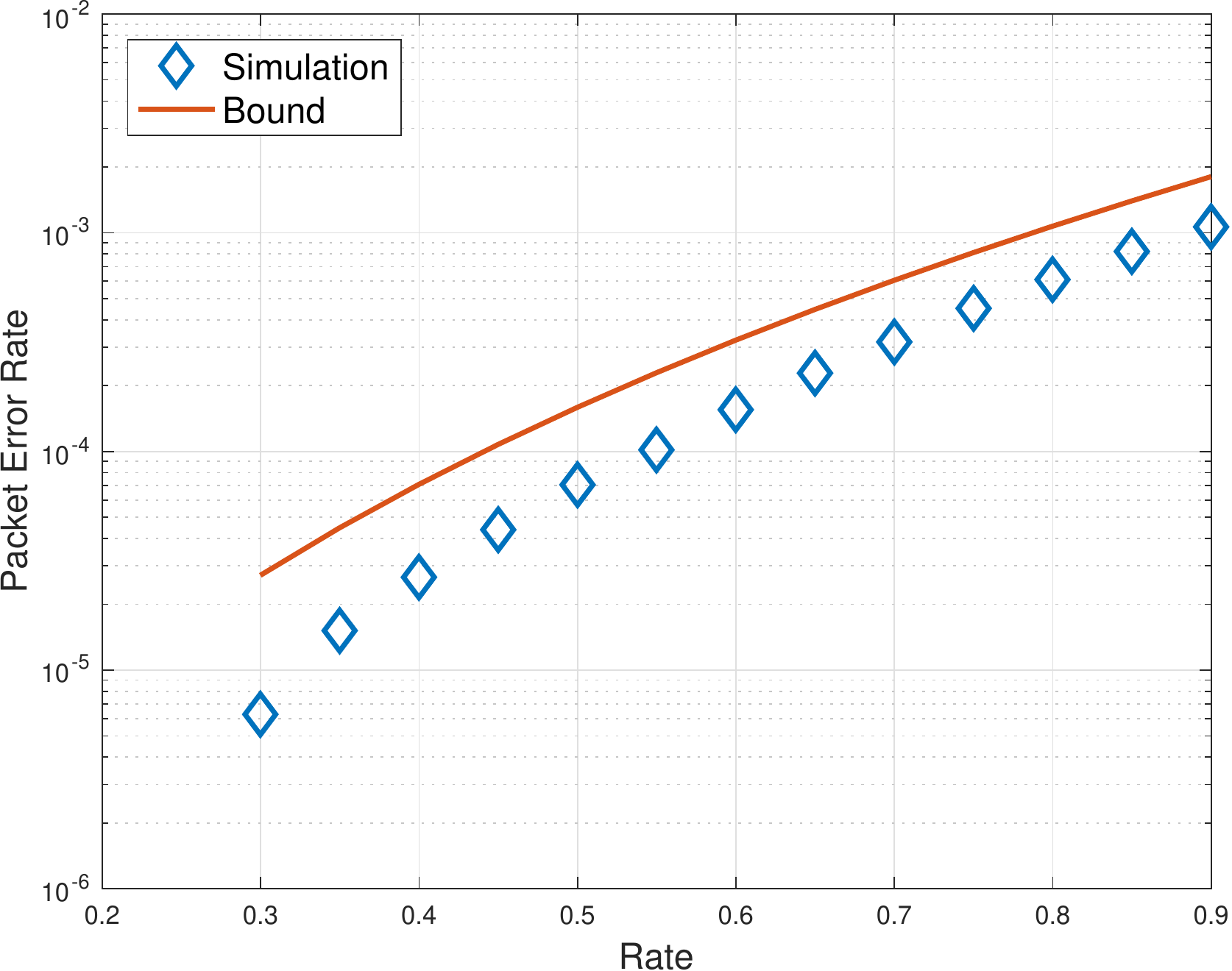}
\end{center}
\caption{Packet error rate as a function of rate, $R$, when
$n = 2^{12}$, $\frac{P}{N_0} = 3$ dB, and $L = 4$.} 
        \label{Fig:sim1}
\end{figure}

Fig.~\ref{Fig:sim1} shows the 
packet error rate as a function of rate, $R$, when
$n = 2^{12}$, $\frac{P}{N_0} = 3$ dB, and $L = 4$.
We can see that the bound becomes tighter as $R$ increases.

\begin{figure}[thb]
\begin{center}
\includegraphics[width=\figwidth]{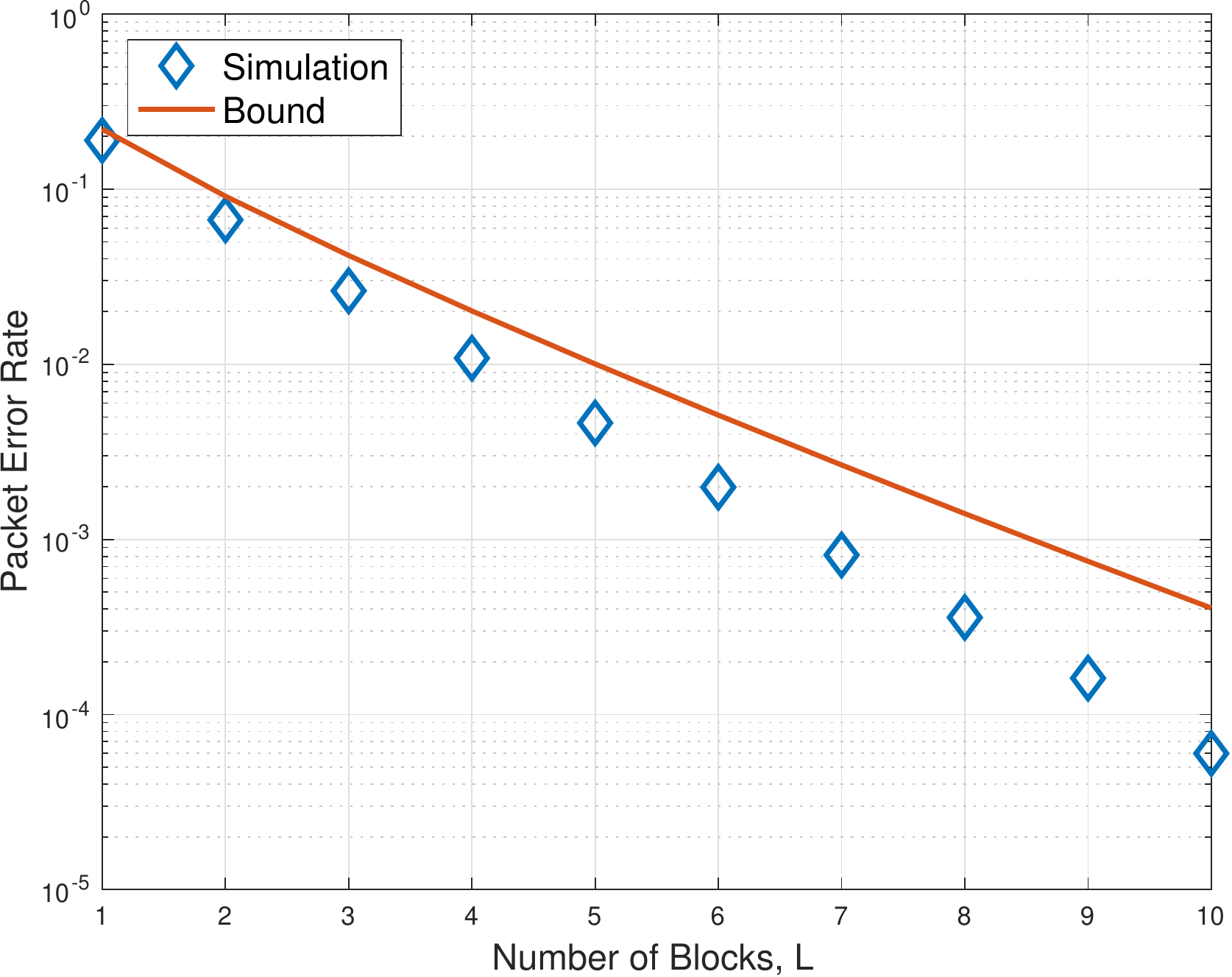}
\end{center}
\caption{Packet error rate as a function of the
number of blocks, $L$, when
$n = 2^{12}$, $R = 1/2$, and $\frac{L P}{N_0} = 3$ dB for all $L
\in \{1, 10\}$.}
        \label{Fig:sim2}
\end{figure}

In Fig.~\ref{Fig:sim2} we show
the packet error rate as a function of the
number of blocks, $L$, when
$n = 2^{12}$, $R = 1/2$, and $\frac{L P}{N_0} = 3$ dB for all $L
\in \{1, 10\}$. Although the total signal power 
remained unchanged when $L$ increases, we can
see that the packet error rate decreases with $L$
thanks to the diversity gain.

\begin{figure}[thb]
\begin{center}
\includegraphics[width=\figwidth]{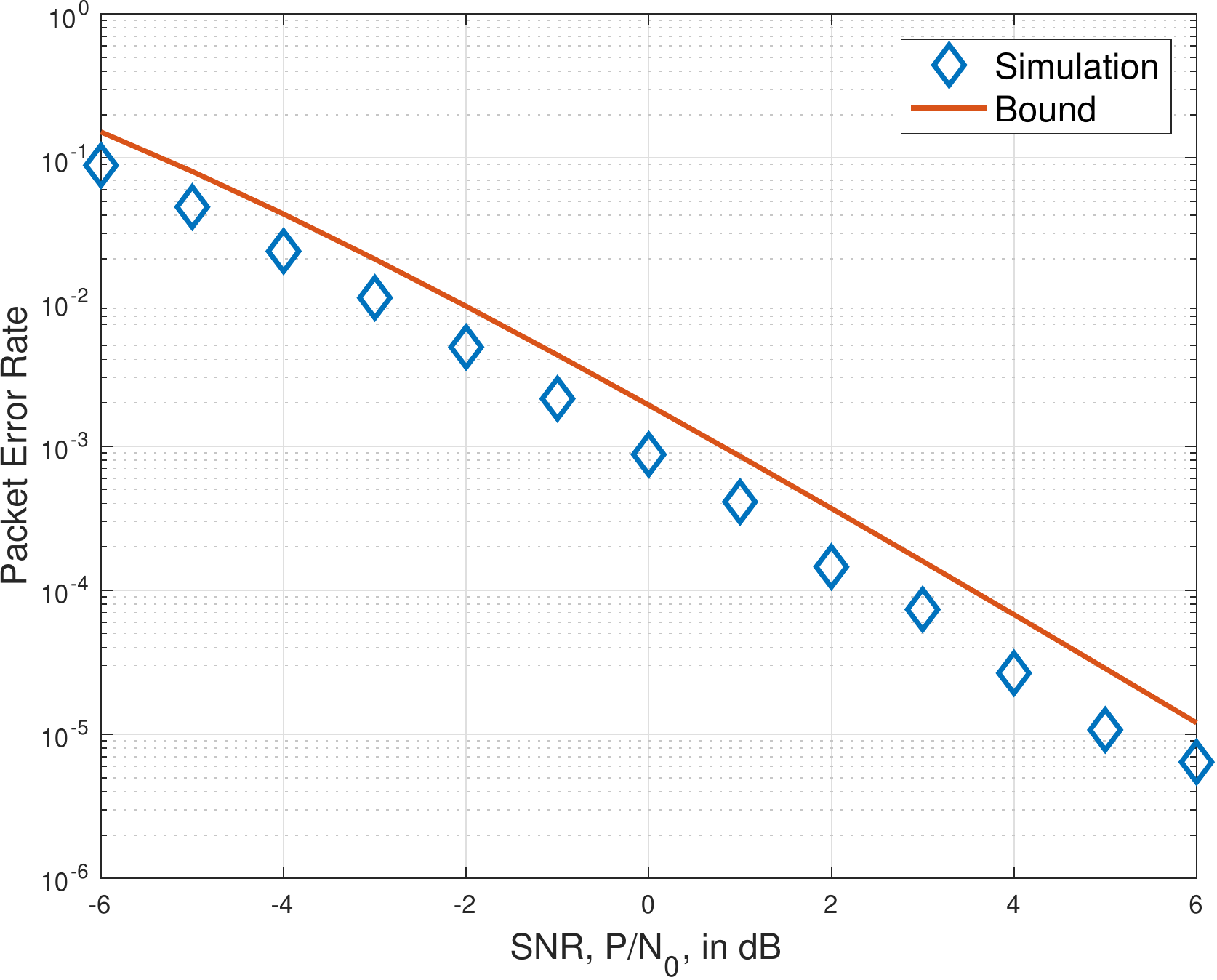}
\end{center}
\caption{Packet error rate as a function of SNR, $\frac{P}{N_0}$,
when $n = 2^{12}$, $R = 1/2$, and $L = 4$.}
        \label{Fig:sim4}
\end{figure}

The impact of the SNR on 
the packet error rate is shown in Fig.~\ref{Fig:sim4}
when $n = 2^{12}$, $R = 1/2$, and $L = 4$.
As expected, 
the packet error rate decreases with the transmit power.
We can also see that the upper-bound is reasonably tight.

\begin{figure}[thb]
\begin{center}
\includegraphics[width=\figwidth]{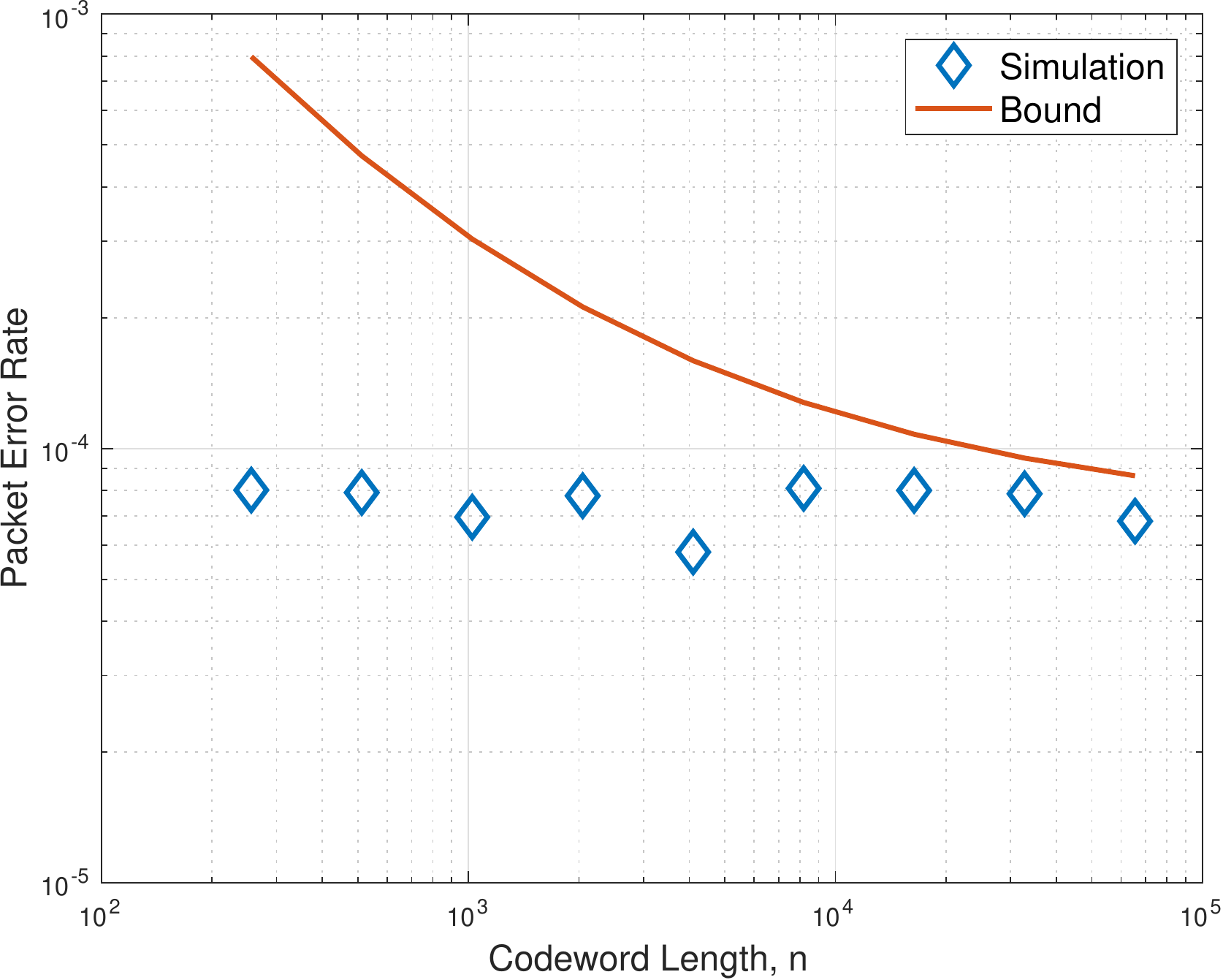}
\end{center}
\caption{Packet error rate as a function of the
codeword length, $n$, when
$R = 1/2$, $\frac{P}{N_0} = 3$ dB, and $L = 4$.}
        \label{Fig:sim3}
\end{figure}

Fig.~\ref{Fig:sim3} shows
the packet error rate as a function of the
codeword length, $n$, when
$R = 1/2$, $\frac{P}{N_0} = 3$ dB, and $L = 4$.
It is shown that the bound becomes tight when
the codeword length is sufficiently long.
If $n$ is small ($\le 1024$), the bound is not tight.

\section{Concluding Remarks}

In this paper, we 
considered multichannel transmissions for URLLC and
derived an upper-bound on the packet error rate 
when finite-length codes are used.
In particular, to take into account fading,
an upper-bound on the packet error rate
was considered with the outage probability.
To find a tight bound, an optimization problem
was formulated and 
a correction term was introduced
for the Chernoff bound on the outage probability.
From simulation results, we found that the derived
upper-bound is reasonably tight and can be
used to determine key parameters in order to guarantee
a sufficiently low packet error rate in URLLC.

\bibliographystyle{ieeetr}
\bibliography{urllc}

\begin{thebibliography}{10}

\bibitem{Soldani18}
D.~{Soldani}, Y.~J. {Guo}, B.~{Barani}, P.~{Mogensen}, C.~{I}, and S.~K. {Das},
  ``{5G} for ultra-reliable low-latency communications,'' {\em IEEE Network},
  vol.~32, pp.~6--7, March 2018.

\bibitem{Bennis18}
M.~{Bennis}, M.~{Debbah}, and H.~V. {Poor}, ``Ultrareliable and low-latency
  wireless communication: Tail, risk, and scale,'' {\em Proceedings of the
  IEEE}, vol.~106, pp.~1834--1853, Oct 2018.

\bibitem{Anand18}
A.~{Anand} and G.~{de Veciana}, ``Resource allocation and {HARQ} optimization
  for {URLLC} traffic in {5G} wireless networks,'' {\em IEEE J. Selected Areas
  in Communications}, vol.~36, pp.~2411--2421, Nov 2018.

\bibitem{Choi19}
J.~{Choi}, ``An effective capacity-based approach to multi-channel low-latency
  wireless communications,'' {\em IEEE Trans. Communications}, vol.~67,
  pp.~2476--2486, March 2019.

\bibitem{Chang95}
C.-S. Chang and J.~A. Thomas, ``Effective bandwidth in high-speed digital
  networks,'' {\em IEEE J. Selected Areas Commun.}, vol.~13, pp.~1091--1100,
  Aug. 1995.

\bibitem{Bockelmann16}
C.~Bockelmann, N.~Pratas, H.~Nikopour, K.~Au, T.~Svensson, C.~Stefanovic,
  P.~Popovski, and A.~Dekorsy, ``Massive machine-type communications in {5G}:
  physical and {MAC}-layer solutions,'' {\em IEEE Communications Magazine},
  vol.~54, pp.~59--65, September 2016.

\bibitem{Malak19}
D.~{Malak}, H.~{Huang}, and J.~G. {Andrews}, ``Throughput maximization for
  delay-sensitive random access communication,'' {\em IEEE Trans. Wireless
  Communications}, vol.~18, pp.~709--723, Jan 2019.

\bibitem{Polyanskiy10IT}
Y.~Polyanskiy, H.~V. Poor, and S.~Verdu, ``Channel coding rate in the finite
  blocklength regime,'' {\em IEEE Trans. Information Theory}, vol.~56,
  pp.~2307--2359, May 2010.

\bibitem{Durisi16}
G.~Durisi, T.~Koch, and P.~Popovski, ``Toward massive, ultrareliable, and
  low-latency wireless communication with short packets,'' {\em Proceedings of
  the IEEE}, vol.~104, pp.~1711--1726, Sept 2016.

\bibitem{Shirvan}
M.~{Shirvanimoghaddam}, M.~S. {Mohammadi}, R.~{Abbas}, A.~{Minja}, C.~{Yue},
  B.~{Matuz}, G.~{Han}, Z.~{Lin}, W.~{Liu}, Y.~{Li}, S.~{Johnson}, and
  B.~{Vucetic}, ``Short block-length codes for ultra-reliable low latency
  communications,'' {\em IEEE Communications Magazine}, vol.~57, pp.~130--137,
  February 2019.

\bibitem{Wolf17}
A.~Wolf, P.~Schulz, D.~Oehmann, M.~Doerpinghaus, and G.~Fettweis, ``On the gain
  of joint decoding for multi-connectivity,'' in {\em GLOBECOM 2017 - 2017 IEEE
  Global Communications Conference}, pp.~1--6, Dec 2017.

\bibitem{Rao18}
J.~{Rao} and S.~{Vrzic}, ``Packet duplication for {URLLC} in {5G} dual
  connectivity architecture,'' in {\em 2018 IEEE Wireless Communications and
  Networking Conference (WCNC)}, pp.~1--6, April 2018.

\bibitem{BiglieriBook}
E.~Biglieri, {\em Coding for Wireless Channels}.
\newblock New York: Springer, 2005.

\bibitem{ChoiJBook2}
J.~Choi, {\em Optimal Combining and Detection}.
\newblock Cambridge University Press, 2010.

\bibitem{ProakisBook}
J.~Proakis, {\em Digital Communications}.
\newblock McGraw-Hill, fourth~ed., 2000.

\bibitem{Tan15}
V.~Y.~F. Tan and M.~Tomamichel, ``The third-order term in the normal
  approximation for the {AWGN} channel,'' {\em IEEE Trans. Information Theory},
  vol.~61, pp.~2430--2438, May 2015.

\bibitem{Mitz05}
M.~Mitzenmacher and E.~Upfal, {\em Probability and Computing: Randomized
  Algorithms and Probability Analysis}.
\newblock Cambridge University Press, 2005.

\bibitem{AbramowitzBook}
M.~Abramowitz and I.~A. Stegun, {\em Handbook of Mathematical Functions with
  Formulas, Graphs, and Mathematical Tables}.
\newblock New York: Dover Publications, 1972.

\bibitem{Simsek17}
M.~Simsek, D.~Zhang, D.~Öhmann, M.~Matthé, and G.~Fettweis, ``On the
  flexibility and autonomy of {5G} wireless networks,'' {\em IEEE Access},
  vol.~5, pp.~22823--22835, 2017.

\end{thebibliography}

\end{document}